# Effects of Defect on Work Function and Energy Alignment of PbI$_2$: Implications for Solar Cell Applications

Hongfei Chen[§], Hejin Yan[§], Yongqing Cai[§]*

[§] Joint Key Laboratory of the Ministry of Education, Institute of Applied Physics and Materials Engineering, University of Macau, Taipa, Macau, China

**ABSTRACT**

Two-dimensional (2D) layered lead iodide (PbI$_2$) is an important precursor and common residual species during the synthesis of lead-halide perovskites. There currently exist some debates and uncertainties about the effect of excess PbI$_2$ on the efficiency and stability of the solar cell with respect to its energy alignment and energetics of defects. Herein, by applying the first-principles calculations, we investigate the energetics, changes of work function and the defective levels associated with the iodine vacancy (V$_I$) and interstitial iodine (I$_I$) defects of monolayer PbI$_2$ (ML-PbI$_2$). We find that the PbI$_2$ has a very low formation energy of V$_I$ of 0.77 and 0.19 eV for dilute and high concentration, respectively, reflecting coalescence tendency of isolated V$_I$, much lower than that of vacancies in other 2D materials like phosphorene. Similar to V$_I$, a low formation energy of I$_I$ of 0.65 eV is found, implying a high population of such defects. Both defects generate in-gap defective levels which are mainly due to the unsaturated chemical bonds of *p*-orbitals of exposed Pb or inserted I. Such rich defective levels allow the V$_I$ and I$_I$ as the reservoir or sinks of electron/hole carriers in PbI$_2$. Our results suggest that the remnant PbI$_2$ in perovskite MAPbI$_3$ (or FAPbI$_3$) would play dual opposite roles in affecting the efficiency of the perovskite: (1) Forming Schottky-type interface with MAPbI$_3$ (or FAPbI$_3$) in which the built-in potential would facilitate the electron-hole separation and prolong the carrier lifetime; (2) Acting as the recombination centers due to the deep defective levels. To promote the efficiency by the Schottky effect, our work reveals that the I$_I$ defect is favored, and to reduce the recombination centers the V$_I$ defect should be suppressed. Our results provide a deep understanding of the effects of defect-engineering in ML-PbI$_2$, which shall be beneficial in improving strategies for the related optoelectronics applications.

**Introduction**

Hybrid organic-inorganic perovskite materials have recently attracted widespread attention in photoelectric fields due to their superior features of high absorption coefficients[1], desirable optical band gaps[2], long carrier lifetime and diffusion length[3]. Specifically, the power conversion efficiency (PCE) of perovskite solar cells (PSCs) has reached ~25.5%[4], while the key to realizing a good performance of PSCs is indispensable with the uniform and well-crystallized perovskite absorbed films[5]. During the one-step reaction, changing the stoichiometry of the precursors $PbI_2$ provides a simple means of varying the chemical potential and the outcome of the reaction, including the defect concentration, doping, and final composition[6]. In two-step synthesis methods, the lead iodide ($PbI_2$) film is deposited and subsequently exposed to solutions of the organic salts thus obtaining the perovskite thin films after thermal annealing[7, 8]. However, the $PbI_2$ in the bottom layer cannot completely interpenetrate with the organic solution in the upper layer, resulting in uncontrollable $PbI_2$ residue. The reaction process in solution proceeds by topotactic nucleation followed by grain growth by dissolution–reconstruction. There is a strong orientational and structural relationship between the final stage of the solution-reacted $MAPbI_3$ product and the initial $PbI_2$ crystallite[9-11]. As the precursor material used to fabricate perovskite thin films[12, 13], $PbI_2$ is critical for efficient PCE due to the $6s^2$ electronic configuration of Pb atoms[14]. More notably, a slight excess of $PbI_2$ in the perovskite film has been observed to be beneficial[15-18]. The remnant $PbI_2$, randomly distribute in the perovskite layer, on the surface or the grain boundary, can suppress charge recombination and reduce the intrinsic halide vacancy concentration thereby passivate trap states[8, 15, 19-22]. It was also theoretically confirmed that the slight excess $PbI_2$ is able to reduce surface cation concentration and the corresponding nonadiabatic coupling decreased by one order of magnitude, hence the charge carrier lifetime extended[23]. However, the excess/unreacted $PbI_2$ will decompose into lead and iodine under illumination, which trigger and accelerate the degradation of PSCs[24].

Actually, $PbI_2$, with layered crystalline structure composed of covalently bonded I–Pb–I atomic planes, is a promising candidate as two-dimensional (2D) semiconductor material owing to its superior optoelectronic properties[25-27]. The 2H phase is the most thermal stable structure for bulk $PbI_2$ at room temperature[28], but the stacking forms of the adjacent $(I-Pb-I)_n$ layers are diverse due to the weak van der Waals interactions[29, 30]. In contrast to other layered semiconductors like $MoS_2$ and $WS_2$, which prefer to n-type behavior, the $PbI_2$ tends to have a p-type feature with tunable bandgap which strongly depends on the thickness[31-33]. As theoretical and experimental researches shown, the band structure of $PbI_2$ would change from a direct bandgap of ~2.38 eV to an indirect bandgap of ~2.5 eV as the thickness decrease from bulk to monolayer[27, 34, 35]. These unique properties trigger many studies on the growth, exfoliation, and applications of the $PbI_2$ nanostructure[34, 36-40] in optoelectronic devices, photodetectors, and photon detection[41-43]. In this work, we prefer to focus on monolayer-$PbI_2$ (ML-$PbI_2$), according to the following reasons: (1) The evidence for the ML-$PbI_2$ has been discovered by electrocrystallisation[44] and the crystalline state of $PbI_2$ has the hexagonal $CdI_2$ type structure. (2) The excess $PbI_2$ species are distributed as nanosheets between the grain boundaries of the perovskite film[20, 45]. (3) In 2D $PbI_2$, low dimensionality provides a sensitive conductive channel for carriers and reduces phonon thermal transport at the same time[46]. (4) Moreover, it is possible to fabricate ML-$PbI_2$ with other 2D materials, such as graphene, by epitaxial growth and layer-control in photovoltaic and thermoelectric systems[39, 47, 48].

Point defects, which are probably to be unintentionally introduced during fabrication and applications, are common in high specific surface area 2D materials. Besides, it is known that defective states, which are situated within the bandgap, affect dramatically carrier mobility and photocatalysis of 2D material. Such defective states can tune band gaps for enhancing light absorption, act as active sites for separating photogenerated carriers, and serve in molecular chemical adsorption with significantly decreased energy barrier of reactions[47, 49-51]. In previous studies, volatile species (for example $I_2$) were found to rapidly decompose perovskites to causes severe degradation[52, 53]. Besides, single-point iodine vacancy ($V_I$) of sublattice sites in ML-$PbI_2$ has also been observed in the previous experiments[48]. Related properties of ML-$PbI_2$ with $V_I$ and nonmetallic atoms doping were theoretically investigated[54]. The results pointed that spin-orbit coupling (SOC) does not alter the electron occupation of the bands near the Fermi level. However, the effect of $V_I$, in particularly its concentration, and interstitial iodine ($I_I$) on ML-$PbI_2$ remains unexplored. Considering the prominent role of $PbI_2$ in organic-inorganic perovskite, further understanding of the defect mechanisms of ML-$PbI_2$ that are predominantly involved with $V_I$ and $I_I$ is thus critically important for practical applications.

In this work, we explored the energetics, changes of work function, and the electronic properties of the pristine ML-PbI$_2$, defective levels associated with the V$_I$ in different concentrations and I$_I$ in two interacted sites by first-principles calculations. Our calculations reveal that the ML-PbI$_2$ has a relatively low formation energy of V$_I$ and I$_I$, implying a high population of such defects under thermodynamic equilibrium conditions. Moreover, the in-gap defective levels in both V$_I$ and I$_I$ allow the reservoir or sinks of electron/hole carriers in PbI$_2$. It is suggested that the remnant PbI$_2$ in perovskite MAPbI$_3$ (or FAPbI$_3$) would play dual opposite roles of the perovskite: (1) Forming Schottky-type interface with MAPbI$_3$ (or FAPbI$_3$) in which the built-in potential would facilitate the electron-hole separation and prolong the carrier lifetime; (2) Acting as the recombination centers due to the deep defective levels. The I$_I$ is favored in order to promote the efficiency by the Schottky effect, and the V$_I$ would be suppressed for reducing the recombination centers.

**Method**

The theoretical calculations were performed using density functional theory (DFT) method as implemented in the Vienna ab initio simulation package (VASP)[55, 56]. The generalized gradient approximation (GGA) of Perdew−Burke−Ernzerhof (PBE) exchange-correlation functional was adopted for geometry optimization and electronic properties[57, 58], and the Heyd–Scuseria–Ernzerhof (HSE06) functional[59] was further employed in the density of state (DOS) and band structure calculations for describing the electronic interactions more precisely. The plane wave basis was applied with a kinetic cutoff energy of 300 eV. The Brillouin zone was sampled with a 2×2×1 and a 1×1×1 Gamma centered Monkhorst-Pack grids for PBE and HSE calculations, respectively. The dispersion correction of DFT-D3 was also considered[60]. All structures were fully relaxed under the convergent criterion of 0.01 eVÅ$^{-1}$ for atomic force. The relaxed lattice constant of PbI$_2$ unit cell is $a = b = 4.593$ Å, which is in good agreement with the experiment result (4.557 Å)[34]. Meanwhile, a larger 6×6×1 supercell (36 Pb atoms and 72 I atoms contained) of ML-PbI$_2$ was also considered. An 18 Å vacuum space was added on the z direction to eliminate the effect of image atoms. The work function $W$ is calculated via the formula:

$$W = E_{vac} - E_F \tag{1}$$

where $E_{vac}$ is the energy in vacuum level, and $E_F$ is the Fermi energy. The band diagram of defect states is aligned through calibrating $E_{vac}$.

Formation energy $E_f$ of defects in ML-PbI$_2$ per single V$_I$ or single I$_I$ was calculated as:

$$E_f = \frac{1}{n}[E_{tot}(Pb_{36}I_{72+n}) - E_{tot}(Pb_{36}I_{72}) - \sum n\mu_I] \qquad (2)$$

where $E_{tot}(Pb_{36}I_{72+n})$ is total energy of the ML-PbI$_2$ supercell with vacancies or interstitials, $E_{tot}(Pb_{36}I_{72})$ is the total energy of prinstic ML-PbI$_2$, $n$ is the number of I atoms that have been added ($n > 0$) to or removed ($n < 0$) from the supercell when the V$_I$ or I$_I$ doping is introduced, and $\mu_I$ is the chemical potentials. The value of the $\mu_I$ of I is defined within the range of values corresponding to Pb-rich or I-rich conditions. Under I-rich condition, the $\mu_I$ is equal to the energy of a I atom in a I$_2$ molecule. Under Pb-rich condition, $\mu_I$ is determined as half of the energy difference between one formula unit of supercell ML-PbI$_2$ and a Pb atom in bulk.

**Results and discussion**

The original 1T phase ML-PbI$_2$ nanosheet, obtained from the 2H-PbI$_2$ bulk[28], has a hexagonal structure with space group $P\bar{3}m1$. Its atomic structure is similar to the 1T-TMDs (MoS$_2$ and WS$_2$), where the atomic plane of Pb is enclosed by two nonaligned I atomic layers to form the ABC stacking pattern, each Pb atom connects with six nearest I atoms forming an octahedral coordination unit [PbI$_6$]$^{4-}$. The optimized lattice constant and thickness of single layer PbI$_2$ are 4.59 Å and 3.78 Å, respectively. The averaged interatomic distance of Pb-I is 3.27 Å, which shows good agreement with the experimental value (3.24 Å)[35, 36].

The defect states including the V$_I$ and I$_I$ are schematically shown in Fig. 1. The insets show the top and side views of the optimized defective states, corresponding to: (a) I$_I$ near the I atom in the basal plane (denoted as I$_{I-I}$), (b) I$_I$ above the Pb atom (denoted as I$_{I-Pb}$), (c) V$_I$. We have considered both the diluted V$_I$ with concentration of 1.38% (denoted as V$_{I-1}$) and more concentrated vacancies with content of 9.72% (V$_{I-2}$) and 19.44% (V$_{I-3}$) for simulating the PbI$_2$ with low, medium and high concentrations, respectively. Such iodine deficiencies are highly likely to exist in remnant PbI$_2$ during the reactions from PbI$_2$ to perovskite PSCs. Although there exist ionic component during the synthesis, the neutral charge state seems relatively stable at a large range of chemical potential[48, 61], so we mainly concentrate on the stability of diluted defects on neutral charge states. For the latter two cases, the vacancies are modelled by removing iodine atoms randomly and uniformly at the top and bottom iodine layers (see more information in Fig. S1 and S2 in supporting information). Since the $PbI_2$ lattice has relatively big void and may uptake some iodine atoms, we have also considered the possibility of the $I_I$ atom located within the sheet of I-Pb-I atomic layer. However, the optimized structure is quite similar to the situation of $I_{I-I}$ owing to the repulsion from hosted atoms. Moreover, the coexistence of vacancy-interstitial (V-I) pairs is also discussed by employing the climbing image nudged elastic band (NEB) method. This kind of Frenkel defects is common in 2D materials and extensively discussed, like in black phosphorus[62]. The energy profiles corresponding to the formation and recombination of V-I pairs are shown in Fig. S3 and S4. We found that the V-I pairs are prone to interact or recombine if concentrations are high.

The formation energies $E_f$ are calculated and listed in Table 1. At finite temperature T, the defect population is exponentially related to $E_f$ according to the Boltzmann distribution. The average equilibrium concentrations defects in ML-$PbI_2$ can be estimated by using $n/N = \exp(-\Delta E/k_B T)$, where $n$ is the number of defects, $N$ is the total number of atomic sites in $PbI_2$, $k_B$ is the Boltzmann constant, and $\Delta E$ is the formation energy with a value of 0.77 eV for $V_{I-1}$, 0.68 eV for $I_{I-I}$ and that of 0.65 eV for $I_{I-Pb}$. The thermodynamic equilibrium concentrations of $V_{I-1}$ (~$10^{-13}$ cm$^{-2}$), $I_{I-I}$ and $I_{I-Pb}$ (~$10^{-12}$ cm$^{-2}$) are low at room temperature, however their concentrations would increase rapidly with increasing temperature. Moreover, high concentrations of defects can also be achieved by irradiation, which is a well-accepted approach for the introduction and control of defects.

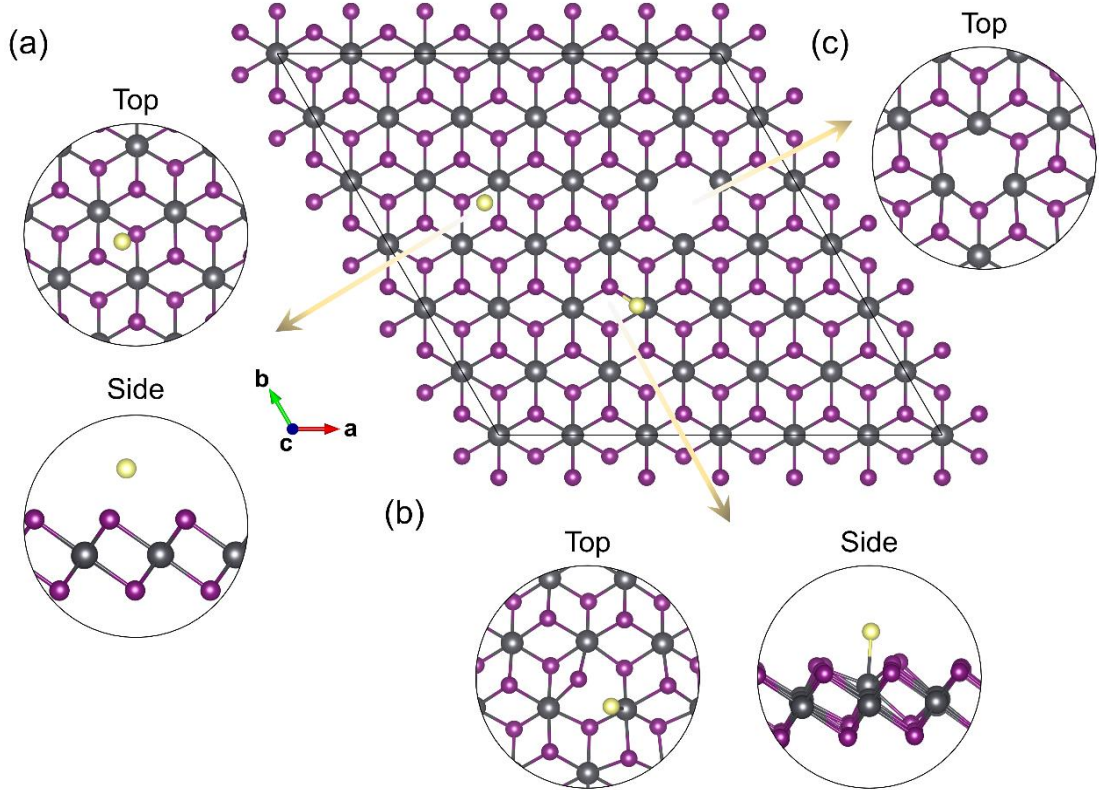

**Figure 1.** Top view of ML-PbI$_2$ 6×6×1 supercell with schematically assembled unrelaxed defective structures of isolated V$_I$ and I$_I$. The dark gray balls represent Pb atoms, the purple balls represent I atoms and the yellow balls represent the interstitial I atoms. The insets show the top and side views of each relaxed configurations of (a) I$_{I-I}$, (b) I$_{I-Pb}$ and (c) V$_{I-1}$.

**Pristine ML-PbI$_2$:**

At first, the electronic properties of the perfect ML-PbI$_2$ and corresponding Pb-I bonding mechanism were investigated. It is well known that the standard PBE functional underestimates the band gap ($E_g$) of monolayer PbI$_2$, and more precise band structure and $E_g$ can be corrected by the hybrid functional. The calculated results of the ML-PbI$_2$ band structure are displayed in Fig. 2a and b. The $E_g$ is 2.55 eV for PBE, which is in good agreement with previously DFT result (2.57 eV)[35, 46]. In addition, this value changes when using the more reliable HSE06 functional with the $E_g$ increases to 3.32 eV. The overall shape of the valence and conduction bands of pristine ML-PbI$_2$ is similar for both PBE and HSE methods of calculation. For ML-PbI$_2$, the conduction band minimum (CBM) is located at Γ point, while the valence band maximum (VBM) is situated at the K point. There presents an energy difference between the direct bandgap at the Γ point and the indirect bandgap (VBM at the K point and CBM at the Γ point) of 0.02 eV in PBE results, while 0.04 eV in HSE results. This direct (bulk PbI$_2$) to indirect (ML-PbI$_2$) transition is mainly originated from the elimination of van der Waals interaction. We find the anti-orbital hybridization between the $5p$ orbitals of I atoms and $5s$ orbitals of Pb atoms plays an essential role in this phenomenon.

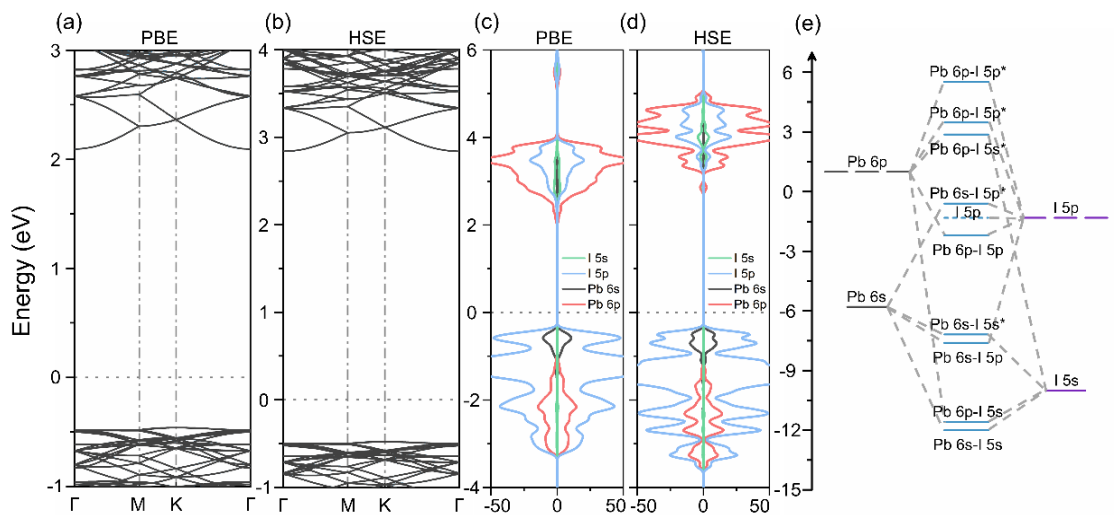

**Figure 2.** Band structure of pristine ML-PbI$_2$ by (a) PBE and (b) HSE methods. PDOS of pristine ML-PbI$_2$ by PBE (c) and HSE (d). In (c) and (d), the Pb-$6s$, Pb-$6p$, I-$5s$ and I-$5p$ states are represented by black, red, green and blue lines, respectively. (e) Diagram of molecular orbital hybridization between Pb and I atoms analyzed from (a). The Fermi level is set at zero, the blue dashed line in (e) represents the nonbonding I $5p$ states.

Figure 2c and d show the projected density of state (PDOS) of pristine ML-PbI$_2$ using PBE and HSE methods, respectively. The number of valence electrons (atomic orbitals) explicitly treated is 14 for Pb ($5d^{10}6s^26p^2$) and 7 for I ($5s^25p^5$). It should be noted that the Pb $5d$ subshell is full and not involved in the bonding thus not analyzed. According to the orbital decomposition analysis of the PBE band structure (Fig. 2a), the Pb $6p$ orbitals dominate the CBM, while the I $5p$ orbitals and Pb $6s$ orbitals mainly contribute to the VBM. Figure 2e shows the orbital hybridization between Pb and I atoms, which well reveals the bonding state of PbI$_2$. The valence bands (VBs) near the Fermi level are composed of Pb ($6s$)-I ($5p$) antibonding states (-1.10 ~ -0.38 eV), I ($5p$) nonbonding states (~ -1.32 eV), and Pb ($6p$)-I ($5p$) bonding states (-3.29 ~ -1.10 eV). Some other bonding or hybridized states are found below the top broad VBs, but too deep and thus less contributing to the physical/chemical processes. Specifically, the Pb (6$s$)-I (5$s$) antibonding states are formed within 0.87 eV (-7.80 ~ -6.93 eV), while in the deeper position of this range the Pb (6$s$)-I (5$p$) hybridized states are detected (The larger energy scale for Fig. 2c shows in Fig. S5). The lowest energy bands are Pb (6$p$)-I (5$s$) and Pb (6$s$)-I (5$s$) weak bonding states (-12.29 ~ -11.32 eV). For the conduction bands (CBs), the weak Pb (6$p$)-I (5$s$) antibonding states and strong Pb (6$p$)-I (5$p$) antibonding states (2.08 eV ~ 4.01 eV) are observed above the Fermi level. We also compare the defective levels of different charged states (-1, 0, +1) of $V_{I-1}$ and $I_{I-Pb}$ defects aligned to vacuum level (Fig. S6) and there are some variations of the positions of the positions.

**Iodine vacancies**

The intrinsic defects, like atomic vacancies are inevitably introduced in PbI$_2$ during the preparation processes. The doping of vacancies has a great influence on the electronic properties, such as enhancing light absorption for photocatalyst, acting as molecular adsorption sites and channels for electrons and holes transfer[63, 64]. Here, we investigated the effect of various $V_I$ ($V_{I-1}$, $V_{I-2}$, $V_{I-3}$) corresponding to the concentrations of 1.38%, 9.72%, and 19.44% on the structural and electronic properties of ML-PbI$_2$. The formation energies $E_f$ are calculated and listed in Table 1. The $E_f$ of all the $V_I$ are positive under both Pb-rich and I-rich conditions implying endothermic processes and additional energy is required for the vacancy production. As expected the $V_I$ under Pb-rich condition has a lower $E_f$ than that under I-rich condition, indicating that $V_I$ is more prone to form under Pb-rich condition. Under dilute doping, the $E_f$ of $V_{I-1}$ ranges from 0.77 eV (Pb-rich condition) to 2.65 eV (I-rich condition). The value is even smaller than those of phosphorus vacancy in phosphorene (1.65 eV), $V_S$ in MoS$_2$ (from 1.22 to 2.25 eV), $V_N$ (from 4.90 to 7.60 eV) and $V_B$ (from 7.50 to 10.20 eV) in $h$-BN[65]. This ultralow $E_f$ in PbI$_2$ suggests a very weak Pb-I bond and facile introduction of isolated $V_I$ in PbI$_2$.

Surprisingly, when more vacancies are introduced in the PbI$_2$ sheet, the $E_f$ can be further reduced. For the medium-level concentrated $V_{I-2}$ and highly concentrated $V_{I-3}$, the $E_f$ decreases from that of the single $V_I$ under both Pb-rich and I-rich conditions: $V_{I-2}$ (from 0.50 to 2.39 eV) and $V_{I-3}$ (from 0.19 to 2.08 eV). This implies that $V_I$ tends to couple with each other and form vacancy clusters.

Figure 1c shows the relaxed structure of single vacancy ($V_{I-1}$). Isolated $V_I$ creates three exposed Pb atoms which exhibit an inward relaxation by moving toward the center of the $V_I$, while the nearest neighbour (NN) iodine atoms show an outward relaxation since the distortion of Pb-I bonds. The loss of an I atom in the $PbI_2$ sheet creates three dangling bonds associated with the unpaired atoms of the three NN Pb atoms. The Pb-I bond length between the first-NN Pb atoms and the next-NN I atoms to the $V_I$ is shortened from 3.27 Å to 3.21 Å. When the concentration increases from $V_{I-1}$, $V_{I-2}$ to $V_{I-3}$, the lattice constants of 6×6×1 $PbI_2$ supercell decrease from 27.56, 27.26 to 25.23 Å, respectively, reflecting a shrinkage of the lattice for structures with significant iodine deficiency.

**Table 1**

The formation energy $E_f$ of $V_I$ and $I_I$ ML-$PbI_2$ under Pb-rich and I-rich conditions. The work function $W$ calculated by the difference of vacuum level energy and Fermi energy.

| System | $E_f$ (eV) | | $W$ (eV) |
| --- | --- | --- | --- |
| | Pb-rich | I-rich | |
| perfect | 0 | 0 | 5.91 |
| $V_{I-1}$ | 0.77 | 2.65 | 5.58 |
| $V_{I-2}$ | 0.50 | 2.39 | 5.76 |
| $V_{I-3}$ | 0.19 | 2.08 | 4.60 |
| $I_{I-I}$ | 2.56 | 0.68 | 4.62 |
| $I_{I-Pb}$ | 2.54 | 0.65 | 4.62 |

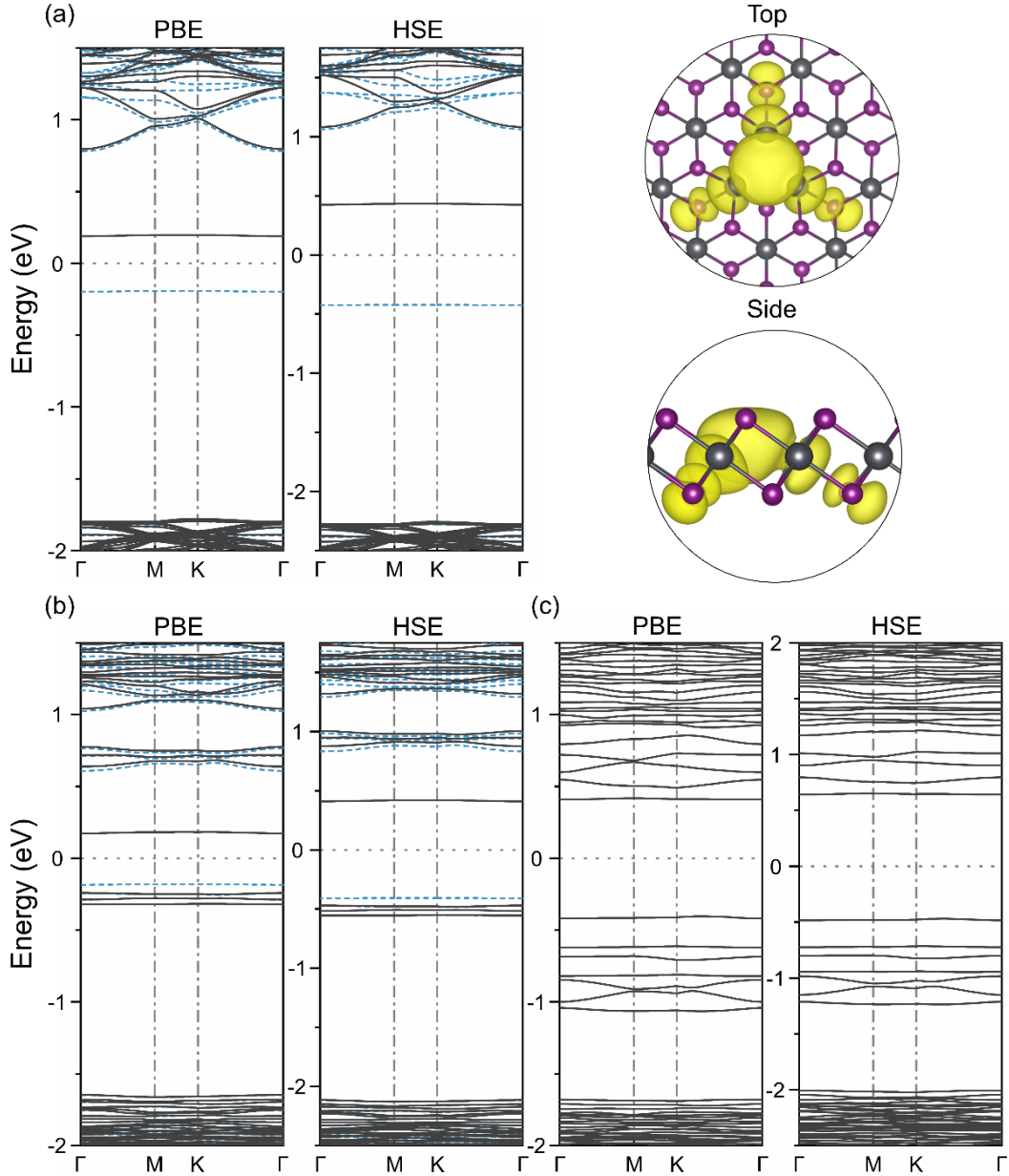

**Figure 3.** Band structures by PBE and HSE methods of ML-PbI$_2$ containing V$_I$. (a) V$_{I-1}$, with V$_I$ concentration of 1.38%, (b) V$_{I-2}$, with V$_I$ concentration of 9.72%, (c) V$_{I-3}$, with V$_I$ concentration of 19.44%. The black solid lines represent the spin-down states and the blue dotted lines represent the spin-up states. The insets are the spatial distribution of localized defective level of V$_{I-1}$ from top and side views.

Figure 3a-c show the band structures of defective ML-PbI$_2$ calculated by PBE and HSE methods. The $E_g$ of V$_{I-1}$ for dilute doping is 2.57 eV (PBE value) and 3.31 eV (HSE value) which are comparable with perfect ML-PbI$_2$. Therefore, the low concentration of V$_I$ has no obvious effect on the $E_g$. However the presence of $V_I$ may lead to non-radiative recombination due to the in-gap defective levels. As illustrated in Fig. 3a, two spin-polarized local impurity levels are generated in the bandgap of ML-PbI$_2$ with the fully occupied spin-up state (blue dotted line) and empty spin-down state (black solid line). Both two flat defect states are deep levels due to their large ionization energy, around 0.60 and 0.98 eV (PBE value) and 0.65 and 1.50 eV (HSE value) below CB edge. These deep energy levels can serve as the recombination center for electrons and holes, thereby shortening the lifetime of non-equilibrium carriers. For instance, the $V_{I-1}$ allows the release of a charged carrier of electron then forms a positive center in ML-PbI$_2$ thus acting as a donor impurity. The insets of Fig. 3 show the charge spatial distributions of $V_{I-1}$ defective levels from the top and side views, where the defective states are highly localized at the three exposed Pb atoms and the $V_I$ core, with slight contributions from the next nearest I atoms from the opposite side.

With increasing the content of iodine vacancies, corresponding to the medium-level $V_{I-2}$ and high-level $V_{I-3}$, the VBs and CBs tend to shift slightly to the Fermi level (Fig. 3b and c). As expected the defective levels associated with $V_I$ become significantly broadened. In particular, for the $V_{I-3}$, those empty defective levels are merged into the CBs, as indicative of a doped degenerating semiconductor, while the lower lying occupied defective bands ranging from -0.42 to -1.06 eV (PBE values) and -0.47 to -1.23 (HSE values) above VB edge. Such rich defective levels could be the sinks or reservoir of photoexcited carriers.

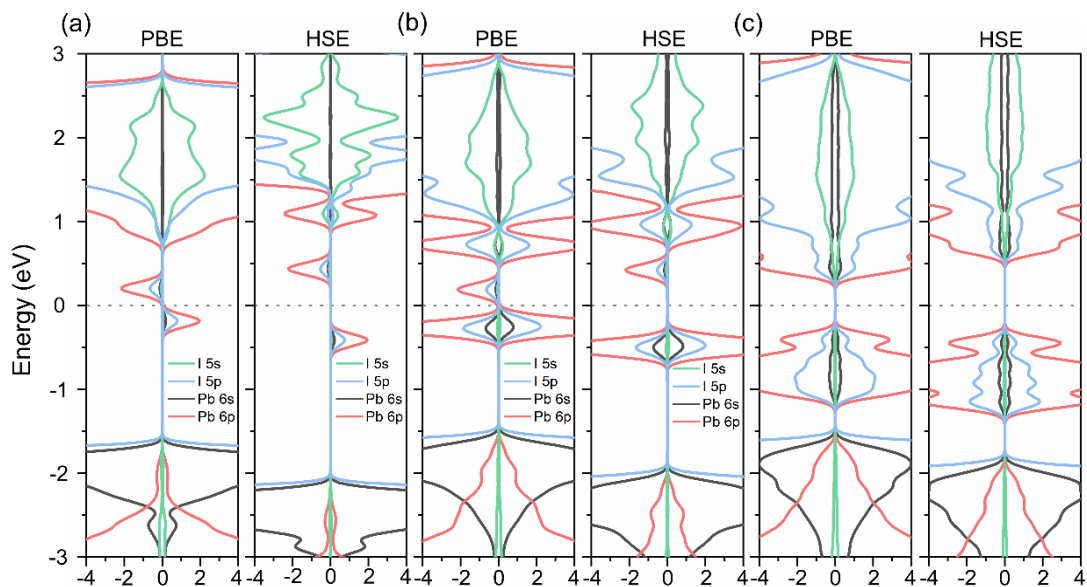

**Figure 4**. Comparison of the PDOS for (a)$V_{I-1}$, (b)$V_{I-2}$, and (c)$V_{I-3}$ by PBE and HSE method, respectively. The Pb 6$s$ orbitals is black lines, Pb 6$p$ orbitals is red lines, I 5$s$ orbitals is green lines, and I 5$p$ orbitals is blue lines.

Figure 4 shows the the PDOS for the evolutions of states with the variation and distributions of vacancies. These impurity levels are mostly composed of 6$p$ states (red lines) corresponding to the NN Pb atoms, with a smaller contribution of 5$p$ states (blue lines) from the NN I atoms and the least for 6$s$ states (black lines) of NN Pb atoms. As seen from Fig. 4a and b, these impurity levels near the Fermi level are localized and spin-dependent. In contrast, as shown in Fig. 4c, the impurity levels become delocalized and have no spin polarization when I-vacancy concentration increases further, developing a band-like extending defective states. One striking feature of highly $V_I$ concentrated $PbI_2$ is that the defective bands, evolved from single $V_I$, keeps a semiconducting behavior, which is different from the metallic behavior of other 2D materials like $MoS_2$[66].

**I interstitial**

Next, we examined the energetics and electronic properties associated with the $I_I$ defect. Here we only consider dilute adsorption of the iodine atom since higher uptake of iodine would lead to similar electronic and structural properties due to the large separation between each two adsorbates. As shown in Fig. 1a and b, there are two different adsorption configurations for interstital iodine: $I_{I-I}$ and $I_{I-Pb}$. For the $I_{I-I}$, the adsorbing I is located and slightly tilted above the I atom in the backbone which triggers minor structural distortion on the PbI$_2$ below. The vertical aligned I-I bond is found with a length of 2.68 Å. For the $I_{I-Pb}$ case, the adsorbing I atom is connected to the Pb atom in the middle layer which induces a strong lattice distortion of PbI$_2$ localized the anchoring site. It is obvious that the presence of interstitial I atom breaks a primitive Pb-I bond and reconstructs a Pb-I bond. A qualitative understanding of the out-of-plane displacement is that the paired electrons of the new Pb-I bond repel the electron on the I atom at the opposite side of PbI$_2$. Surprisingly, the strong bond reorganization in this $I_{I-Pb}$ has a similar $E_f$ with that of surface adsorption in $I_{I-I}$. This implies again a weak bonding energy of Pb-I and explains the facile fabrication of PSCs starting from PbI$_2$ precursor.

The $E_f$ of $I_{I-I}$ and $I_{I-Pb}$ was listed in Table 1 which is similar for these two different configurations. For $I_{I-I}$, the $E_f$ is 2.56 eV (I-poor) and 0.68 eV (I-rich), and for $I_{I-Pb}$ the values are slightly lower with 2.54 eV (I-poor) and 0.65 eV (I-rich). Therefore, compare with the Pb-rich environment, the $I_I$ is more prone to form in I-rich condition.

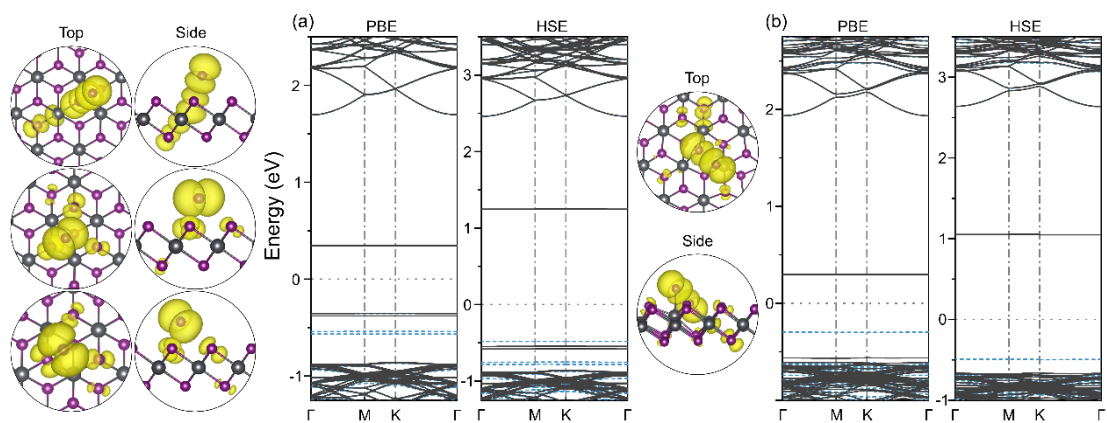

**Figure 5**. Band structures of (a) $I_{I-I}$ and (b) $I_{I-Pb}$. The black solid lines represent the spin-down states and the blue dotted lines represent the spin-up states. The insets on the left side of the band structure show the spatial distribution of the in-gap defective states from top and side views. For the $I_{I-I}$, three defective charge plots are located from top to bottom according to the sequence of reducing energy level.

Figure 5 shows the band structures of $I_{I-I}$ and $I_{I-Pb}$ defective system. The calculated $E_g$ of $I_{I-I}$ are 2.55 eV for PBE and 3.30 eV for HSE whereas it are slightly different for $I_{Pb-I}$: 2.50 eV for PBE and 3.29 eV for HSE. Although these values of $E_g$ are similar to the pristine ML-PbI$_2$ (2.55 eV and 3.32 eV), the interstitial defects induces new characteristics with the introduced defect levels. As shown in Fig. 5a, for I$_{I-I}$ system, six spin-polarized localized impurity levels are created in the bandgap of ML-PbI$_2$. The impurity levels are quite close to the VBs (locating at 0.10, 0.30 and 0.37 eV above the VB edge by HSE), indicating that the interstitial iodine gives rise to shallow acceptor levels. The iodine atom could take the electrons and become a negative center, thus acting as an acceptor impurity, which provides holes carriers in ML-PbI$_2$. Besides, we also notice one defect level is 2.11 eV from VBs by HSE and thus as a deep level, indicated that the I$_{I-I}$ will induce recombination center in ML-PbI$_2$ as well. As shown in the inset of Fig. 5, the defective state mainly consists of $5p$ states of I atoms. These shallow impurity levels are dominated by $5p_x$ and $5p_y$ orbitals of I$_{I-I}$ while the deep level relies on the $5p_z$ orbitals.

As for I$_{I-Pb}$ system, the interstitial iodine atom produces two defect levels, one is an occupied and spin-up state with a shallow nature (0.16 eV above VB by HSE) and another is empty and spin down state with a relatively deep location (1.70 eV above VB by HSE), as shown in Fig. 5b. The real space representation of the state shows that it is mainly contributed by the $5p$ orbital coupling between adsorbed I atom and the I atom from the broken Pb-I bond.

Figure 6 shows the PDOS of I$_{I-I}$ and I$_{Pb-I}$ systems. Consistent with the real space distribution as shown in insets of Fig. 5, the I $5p$ states dominate the localized defective levels whereas the interlayer Pb atom almost have no contribution. Both two interstitial adsorptions induce shallow (near the band edge) defect energy states, which will produce relatively slow electron–hole recombination kinetics. This can be attributed to the charge defect-tolerance of PbI$_2$, which generally form Pb ($6s$)-I ($5p$) and Pb ($6p$)-I ($5p$) antibonding molecular orbitals. Analogously, lead-halide perovskite was found as a defect-tolerant semiconductor with lacking of bonding–antibonding interaction between the conduction bands and valence bands[67-70].

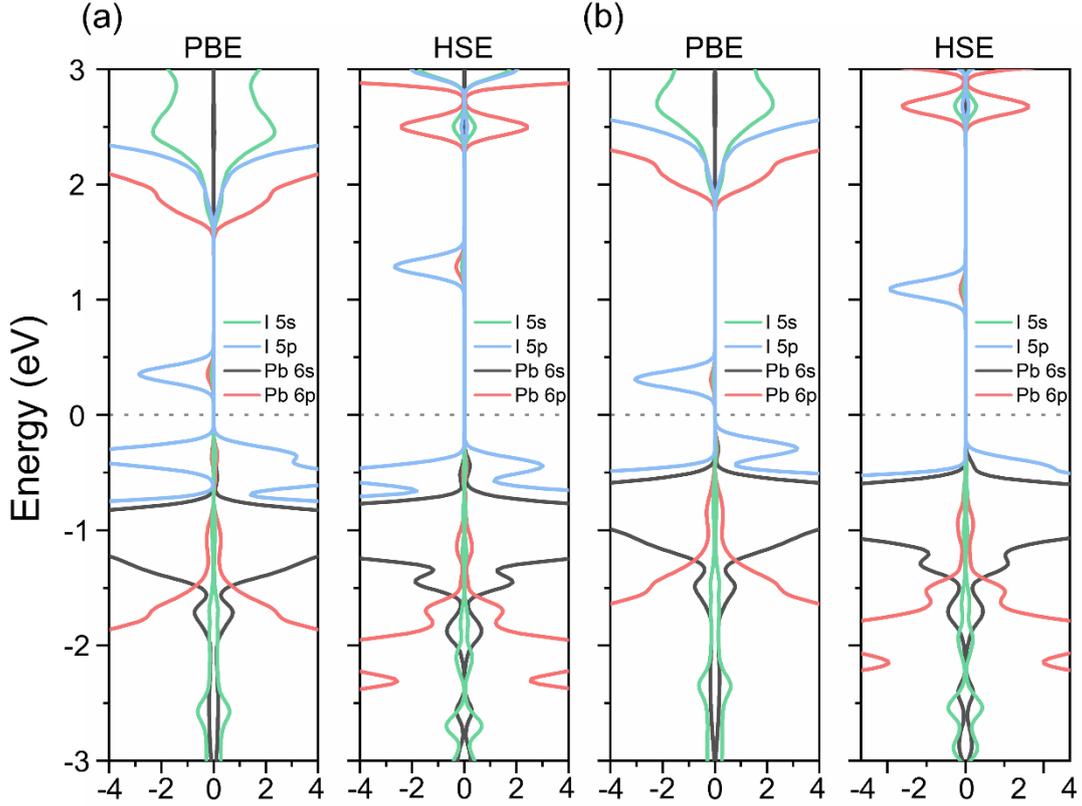

**Figure 6.** Comparison of the PDOS for (a) $I_{I-I}$ and (b) $I_{I-Pb}$. The Pb 6s orbitals is black lines, Pb 6p orbitals is red lines, I 5s orbitals is green lines, and I 5p orbitals is blue lines.

**Band alignment and implications for solar cell application**

Finally we would like to analyze the band alignment between perfect and defective $PbI_2$ together with other common PSCs materials like the $MAPbI_3$, $FAPbI_3$ and electron hole transport materials. Figure 7 depicts the alignment of band edges, levels of point defects in $PbI_2$ and other CBM/VBM of $FAPbI_3$, $MAPbI_3$, $TiO_2$, $SnO_2$, and ZnO from literatures.[71, 72]. Figure 8 shows some possible interfaces formed between $PbI_2$ and $MAPbI_3$, which have the potential to create a Schottky junction between p-type $PbI_2$ and perovskite $MAPbI_3$. The values of work function $W$ of the perfect and defective ML-$PbI_2$ system are listed in Table 1. The work function of perfect $PbI_2$ is 5.91 eV, which drops to 5.58 and 4.60 eV for dilute and heavy doping of vacancies, respectively. Surprisingly, the minor adsorption of interstitial iodine would reduce the $W$ to 4.62 eV.

According to our calculations, the $I_V$ and $I_I$ have a very low formation energy and associated with in-gap defective levels. This means a high density of such defects is highly probable. Our calculated band structure shows the significant broadening of the defective states which could be a band-like extending nature. Owing to its relatively big intrinsic band gap of $PbI_2$, the Fermi level is less likely to cross such defective band-like states, thus the $PbI_2$ still remains a semiconducting state, as shown by our calculations of the high-content vacancy case of $V_{I-3}$. In principle, such rich defective levels allow the reservoir or sinks of electron/hole carriers in $PbI_2$. The $PbI_2$ itself cannot be directly used for photo-excited carriers generation due to its poor thermodynamic stability reflected by the weak Pb-I bond. However, as an important precursor for $MAPbI_3$ and $FAPbI_3$, remnant tiny $PbI_2$ is possible[19, 20] and plays some roles in affecting the efficiency of the perovskite material. In real PSC materials, the $PbI_2$ could play dual roles: (1) Forming Schottky-type interface with $MAPbI_3$ (or $FAPbI_3$): Since the $PbI_2$ is p-type and according to our energy alignment analysis in Fig. 7, the interface would be Schottky type. Figure 8a shows the schematic diagram of Schottky junction between p-type $PbI_2$ and perovskite $MAPbI_3$. The highly dense low-lying defective states in $PbI_2$, mainly from the $I_I$, would raise the Fermi level and reduce the Schottky barrier. The leakage of the electrons from the $FAPbI_3$ or $MAPbI_3$ into the $PbI_2$ would generate the built-in potential across the interface. Such built-in potential would help to facilitate the electron-hole separation and promote the carrier lifetime. (2) Acting as traps or recombination center: As shown in Fig. 7, the high-lying defective states associated with the $V_I$ locate below the CBM of $FAPbI_3$ and $MAPbI_3$ which are empty and allows carriers trapping or recombination. This would reduce the carrier lifetime in the host $FAPbI_3$ or $MAPbI_3$. Therefore, $PbI_2$ would play both negative and positive roles in PSCs, which have a good agreement with previous works[20, 21, 24]. To promote the efficiency by the Schottky effect, the $I_I$ defect is favored, and to reduce the recombination centers the $V_I$ defect would be suppressed.

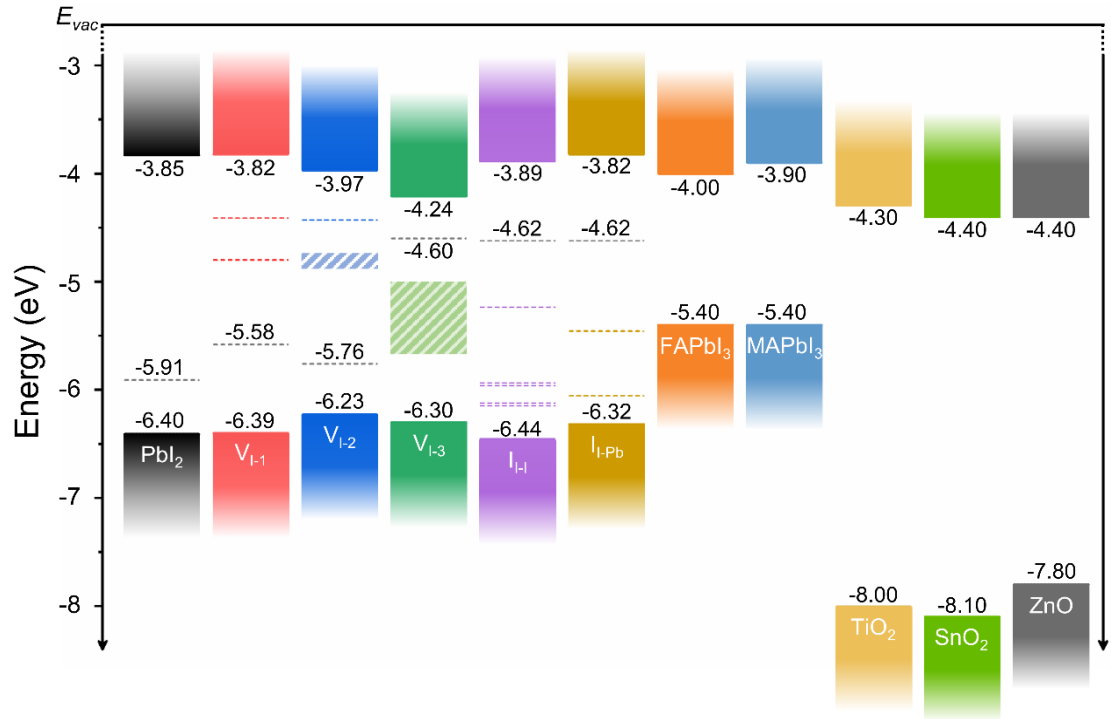

**Figure 7**. Band alignment of pristine and defective ML-PbI$_2$ according to PBE results, compared with some other materials available for PSCs. The positions of the VBM and CBM are labeled, and the defective states in the bandgap are represented with dashed lines and defective bands (for V$_{I-2}$ and V$_{I-3}$ cases only in the light blue and light green slashes, respectively). The levels of work function (Fermi levels relative to the $E_{vac}$) are plotted with dark gray dashed lines.

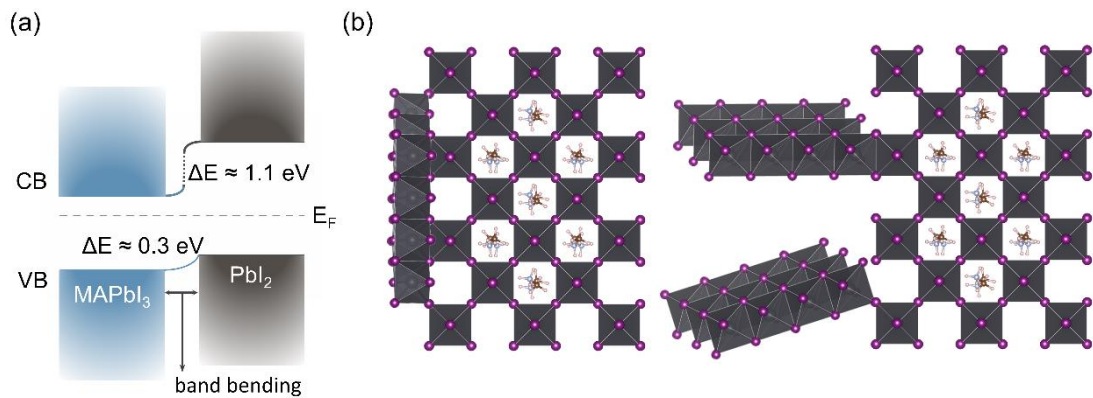

**Figure 8.** (a) Schematic diagram of Schottky junction between p-type PbI$_2$ and perovskite MAPbI$_3$. (b) Artificial illustration of interfaces formed between PbI$_2$ and MAPbI$_3$.

**Conclusion**

In summary, we have investigated the defective states in the PbI$_2$ for the iodine vacancies and interstitial defects by using the first-principles study. Under Pb-rich and moderate growth condition, the iodine vacancy is the dominant defect, whereas under I-rich condition, the iodine interstitial becomes more popular. A noticeable trend was the steady decrease in bandgap with the growth of iodine vacancy, due to the consistent increase in the number of defective levels. For iodine vacancy, it introduces deep levels and forms the carrier recombination center. These defective states act as trapping and scattering centers for conducting carriers and decrease their electronic mobility. For iodine interstitial, shallow trap states and deep levels are found simultaneously, which can be attributed to the charge defect-tolerance analogized to lead-halide perovskite. The low-lying defective levels in remnant PbI$_2$ would allow the formation of Schottky-type junction with MAPbI$_3$ (or FAPbI$_3$) which can prolong the lifetime of carriers in the perovskite due to the built-in potential. Our work sheds some lights on the dual roles of the PbI$_2$ and a proper defect engineering of such residual PbI$_2$ phase would be beneficial for the promoted efficiency of perovskite solar cells.

## ASSOCIATED CONTENT

**Supporting Information**

Relaxed atomic structures of medium and high content of iodine vacancies. This material is available free of charge via the Internet at http://pubs.acs.org.

## AUTHOR INFORMATION

**Corresponding Author**

yongqingcai@um.edu.mo

**Notes**

The authors declare no competing financial interests.

## ACKNOWLEDGMENT

This work is supported by the University of Macau (SRG2019-00179-IAPME) and the Science and Technology Development Fund from Macau SAR (FDCT-0163/2019/A3), the Natural Science Foundation of China (Grant 22022309) and Natural Science Foundation of Guangdong Province, China (2021A1515010024). This work was performed in part at the High Performance Computing Cluster (HPCC) which is supported by Information and Communication Technology Office (ICTO) of the University of Macau.